\documentclass[prl,twocolumn,lengthcheck,superscriptaddress,letterpaper,nofootinbib,nopreprintnumbers,showpacs]{revtex4-1}

\usepackage{color}
\usepackage{graphicx}
\usepackage{float}
\usepackage{amsmath}
\usepackage{acronym}
\usepackage{multirow}
\usepackage{verbatim}
\usepackage{algcompatible}
\usepackage[size=small]{caption}
\usepackage{etoolbox}

\usepackage{color,ulem}

\newcommand{\beq}{\begin{equation}}
\newcommand{\eeq}{\end{equation}}
\newcommand{\bea}{\begin{eqnarray}}
\newcommand{\eea}{\end{eqnarray}}
\newcommand{\ba}{\begin{array}}
\newcommand{\ea}{\end{array}}

\begin{document}

\title{%
Rapidly evaluating the compact binary likelihood function via interpolation
}

\pacs{04.30.-w, 04.80.Nn
}
\author{R. J. E. Smith}
\affiliation{School of Physics and Astronomy, University of Birmingham, Edgbaston, Birmingham B15 2TT, UK}

\author{C. Hanna}
\affiliation{Perimeter Institute for Theoretical Physics, Waterloo, Ontario N2L 2Y5, Canada}

\author{I. Mandel}
\affiliation{School of Physics and Astronomy, University of Birmingham, Edgbaston, Birmingham B15 2TT, UK}

\author{A. Vecchio}
\affiliation{School of Physics and Astronomy, University of Birmingham, Edgbaston, Birmingham B15 2TT, UK}
\date\today

\begin{abstract}
Bayesian parameter estimation 
on gravitational waves from compact binary coalescences (CBCs) typically requires millions of template waveform computations at different values of the parameters describing the binary.
Sampling techniques such as Markov chain Monte Carlo and nested sampling evaluate likelihoods, and hence compute template waveforms, serially; thus, the total computational time of the analysis scales linearly with that of template generation. Here we address the issue of rapidly computing the likelihood function of CBC sources with non-spinning components. We show how to efficiently compute the continuous likelihood function on the three-dimensional subspace of parameters on which it has a non-trivial dependence -- the chirp mass, symmetric mass ratio and coalescence time -- via interpolation. Subsequently, sampling this interpolated likelihood function is a significantly cheaper computational process than directly evaluating the likelihood; we report improvements in computational time of two to three orders of magnitude while keeping likelihoods accurate to $\lesssim 0.025\,\%$.  Generating the interpolant of the likelihood function over a significant portion of the CBC mass space is computationally expensive but highly parallelizable, so the wall time can be very small relative to the time of a full parameter-estimation analysis.

\end{abstract}

\keywords{}

\maketitle

\textit{Introduction---}
The direct detection of gravitational waves will initiate an entirely new kind of astronomy, offering an unprecedented probe of relativistic astrophysics and strong-field gravity. Ground-based gravitational-wave interferometers LIGO \cite{iLIGO} and Virgo \cite{iVirgo}  are undergoing upgrades to their second generation designs, Advanced LIGO/Virgo (aLIGO/AdV), and are expected to be operational around 2015 \cite{aligo, aVirgo}. In addition two new detectors in India and Japan, LIGO India \cite{ligoindia} and KAGRA \cite{KAGRA} respectively, are expected to be operational around 2020. These advanced instruments will be an order of magnitude more sensitive than their predecessors \cite{NoiseCurves} and are expected to usher in routine detections of gravitational waves \cite{LSCVrates:2010}. The coalescence of compact binaries, consisting of neutron stars and/or black holes, are prime targets for gravitational-wave observatories, with realistic estimates of detection rates $\sim 1 - 100\,\mathrm{yr}^{-1}$ \cite{LSCVrates:2010}. Estimating the parameters of CBC sources -- e.g. their masses, spins and sky location -- is a crucial aspect of gravitational-wave astronomy, but remains challenging in practice.

%There are numerous experimental, theoretical, and data analysis challenges associated with detecting, and estimating the parameters of, compact binary coalescence (CBC) sources. Here we focus on addressing the computational time of performing Bayesian parameter estimation on gravitational-wave signals from CBC sources buried in a noisy data stream. CBC sources with non-spinning components are characterized by nine parameters, whereas binaries with spinning components are characterized by fifteen. In order to explore the probability distributions of these parameters in such a high dimensional space one requires sophisticated stochastic Bayesian inference techniques \cite{VV:2010, Raymond:2009, VDS:2008} which preferentially sample the parameter space in regions of high likelihood. Examples of commonly used stochastic sampling algorithms for CBC parameter estimation include Markov Chain Monte Carlo (MCMC) \cite{Raymond:2009, VDS:2008} and Nested Sampling \cite{VV:2010}, both of which sample the parameter space in a Markovian way.

The goal of Bayesian parameter estimation is to compute the posterior probability density function (PDF) of a set of parameters, $\vec{\theta}$, which underlie a model assumed to describe a data set $d$. The PDF is related to the likelihood function and prior probability via Bayes' theorem:
\beq
p(\vec{\theta}| d) = \frac{\mathcal{P}( \vec{\theta})\ \mathcal{L}(d | \vec{\theta}) } { p(d) },\
\label{eq:bayes}
 \eeq
where $\mathcal{L}(d | \vec{\theta} )$ is the likelihood and $\mathcal{P}( \vec{\theta} )$ is the prior probability which encodes our \textit{a priori} belief in the distribution of $\vec{\theta}$. The quantity in the denominator, $p(d)$, is known as the ``evidence''. Computing~(\ref{eq:bayes}) requires evaluating the likelihood. 
%, given by
%\beq
%(a | b) = 4\,\Re\,\int^{f_{\mathrm{max}}}_{f_{\mathrm{min}}}\, df\, \frac{\tilde{a}(f)\tilde{b}^{*}(f)}{S_{n}(f)}\,.
%\label{eq:innerprod}
%\eeq

For binaries with non-spinning components $\vec{\theta}$ is nine dimensional. Exploring such a high dimensional space requires sophisticated stochastic Bayesian inference techniques \cite{VV:2010, Raymond:2009, VDS:2008} which preferentially sample the parameter space in regions of high posterior probability. The bulk of the computational cost of evaluating the likelihood function comes from computing template waveforms.
Analyses on first-generation interferometer data require computing $\mathcal{O}(10^{6})$ such waveforms \cite{Raymond:2009, Smith:2012}.  Sampling techniques such as Markov chain Monte Carlo (MCMC) \cite{Raymond:2009, VDS:2008} and nested sampling \cite{Skilling:2004,VV:2010} evaluate likelihoods, and hence compute template waveforms, serially. Thus the total computational time to fully sample the parameter space scales linearly with the total time spent generating template waveforms.  It can take hours to weeks to analyse a single stretch of data of a few seconds in duration, depending on the choice of the template waveform family. This problem will be exacerbated when analysing second-generation interferometer data as the waveforms will be forty times longer in duration if the starting frequency $f_{\mathrm{min}}$ is changed from $40$\,Hz to $10$\,Hz. 
%This is true whether considering binaries with non-spinning components or spinning components, though the latter case is generally more challenging. 

For binaries with non-spinning components, the frequency-domain waveform $\tilde{h}(\vec{\theta}\, ; f)$ has the schematic form
\beq
\tilde{h}(\vec{\theta}\, ; f) = \sum_{\mu = +\,,\times}A_{\mu}(\vec{\theta}_{L}) \tilde{h}_{0}(\mathcal{M}\,,\eta ;f)e^{2\pi ift_{c}}\,,
\eeq
where $A_{+\,,\times}$ denotes the (scalar) amplitudes of the ``plus-'' and ``cross-'' polarization states of the waveform. In general $\tilde{h}_{0}$ depends on the waveform family being used and can be computed by Fourier transforming the associated time-domain representation of the waveform family. The parameters which describe the binary are the chirp mass and symmetric mass-ratio, $\mathcal{M}$ and $\eta$, the time at coalescence $t_c$ and a set of parameters which describe the location and orientation of the binary $\vec{\theta_{L}}$.

Evaluating the likelihood function on the three-dimensional subspace of parameters $(\mathcal{M}\,, \eta\,, t_c)$ represents the largest computational burden to parameter estimation on gravitational waves from CBC sources with non-spinning components because the likelihood function depends non-trivially on these parameters, and so requires a new waveform evaluation.   In  \cite{Smith:2012}, we considered interpolation between waveforms over the mass parameter space as a way to reduce computational cost.  Here, we 
%Often waveforms are so similar that one may ask if an interpolation scheme is a feasible way to reduce the cost.  We considered this in \cite{Smith:2012}.  Here we}
%Hence, if one could rapidly sample the likelihood function in the  $\mathcal{M}\,, \eta$ and $t_c$ subspace one could in principle reduce the computational time of sampling the likelihood function over the full parameter space. 
demonstrate that the evaluation of an interpolated likelihood function over the $(\mathcal{M}\,, \eta\,, t_c)$ subspace is a much faster computational procedure than the standard calculation of the likelihood~(\ref{eq:logL_expanded}) by using either full or interpolated waveforms. 
%Using interpolated likelihood functions can be a step forward over using interpolated template waveforms for parameter estimation, which we considered in \cite{Smith:2012}.  \citet{Smith:2012} used singular value decomposition (SVD) to represent template waveforms as combinations of basis templates and ``reconstruction coefficients''. The latter can be interpolated which allows template waveforms to be reconstructed.
For the purposes of parameter estimation, one is not interested in template waveforms per se, but rather in the posterior probability distributions of the underlying parameters of the template waveforms that are assumed to describe the data. By directly using interpolated likelihood functions, one effectively bypasses template waveform generation during the sequential steps of an MCMC. This likelihood-interpolation technique is robust and could, in principle, be generalized to arbitrary template waveform families, in particular those that describe CBCs with spinning components.

\textit{Directly interpolating the likelihood function---} 
We wish to generate a representation of the likelihood function over the continuous $\mathcal{M}, \eta$ and $t_c$ subspace. To achieve this we will interpolate the likelihood function over $\mathcal{M}, \eta$ and $t_c$. The likelihood function that describes the probability of observing a data stream $d = h +  n$ containing a given gravitational-wave signal $h(\vec{\theta};\, t)$ and Gaussian and stationary noise $n(t)$ is \cite{VV:2010}
\beq
\log\,\mathcal{L}(d|\vec{\theta}) = (d|h(\vec{\theta})) -\frac{1}{2}\left[ \big(h(\vec{\theta})|h(\vec{\theta})) + (d|d)\right]\,,
\label{eq:logL_expanded}
\eeq
where $(a | b)$ is the usual noise-weighted inner product \cite{Owen:96}. 
%To elucidate the dependence of the likelihood function on the binary parameters we find it useful to 
We define the complex-valued time-series corresponding to the inner product between two time series $a(t)$ and $b(t)$ as one is shifted by an amount $t_c$ with respect to the other:
\beq
z[a\,,b](t_c):=4\,\int_{f_{\mathrm{min}}}^{f_{\mathrm{max}}}\,df\,\frac{\tilde{a}(f)\tilde{b}^{*}(f)}{S_n(f)}\,e^{-2\pi ift_{c}}\,,
\eeq
In the above, $\tilde{a}(f)$ is the Fourier transform of $a(t)$ and $S_{n}(f)$ is the detector noise power spectral density (PSD). The limits of integration are in general specified by the bandwidth over which an analysis is being conducted. In terms of $z(t_c)$ the inner products in~(\ref{eq:logL_expanded}) are succinctly expressed as
\bea
&&(h(\vec{\theta})|h(\vec{\theta})) = \Re\mathcal{A}(\vec{\theta}_{L})z[h_0(\mathcal{M}, \eta),h_0(\mathcal{M}, \eta)](0),\\
&&(d|h(\vec{\theta})) = \Re\mathcal{B}(\vec{\theta}_{L})z[d\,,h_0(\mathcal{M}, \eta)](t_c)\,,
\eea
and $\mathcal{A}(\vec{\theta}_{L})$ and $\mathcal{B}(\vec{\theta}_{L})$ are known projections which contain the $\vec{\theta}_{L}$ dependence in the likelihood function.

We have previously interpolated template waveforms over the mass parameters \cite{Cannon:2012, Smith:2012}, and here we show that the same technique can be applied to interpolating the time series $z[d\,,h_0](t_c)$. The interpolation of $z[d\,,h_0](t_c)$ is based on the SVD of a set of (discretely sampled) time series distributed on a two-dimensional grid. In this case the two-dimensional grid spans $\mathcal{M}$ and $\eta$ and the time parameter is $t_c$. We use the notation $\vec{z}\,[d\,,h_0]$ to describe the discretely sampled $z[d\,,h_0](t_c)$.
% If $\vec{z}\,[d\,,h_0]$ is sampled at a rate $\Delta\,t$ then the components of $\vec{z}\,[d\,,h_0]$ at discrete times $t_{i} = i\Delta\,t$ are $\vec{z}(t_{i}):=z_i[d\,,h_0]$. 
Recall that the SVD of the discretely sampled time series $\vec{z}\,[d\,,h_0]$ allows it to be written as a linear superposition of orthonormal basis vectors $\vec{u}_{\mu}$ and projection coefficients $M_{\mu}$ \cite{Cannon:2010}: 
\beq
\vec{z}\,[d\,,h_0(\mathcal{M}, \eta)] = \sum_{\mu}\, M_{\mu}(\mathcal{M}\,,\eta)\,\vec{u}_{\mu}\,.
\eeq
The coefficients $M_{\mu}$ can be interpolated over $\mathcal{M}$ and $\eta$ and we follow the method in \cite{Cannon:2012} which uses Chebyshev polynomials of the first kind. 

Interpolation of $z[h_0\,,h_0](0)$ over $\mathcal{M}$ and $\eta$ is simple as it is scalar valued and we again use Chebyshev polynomials of the first kind. Below we provide an example of the interpolation technique outlined here.

\textit{Likelihood interpolation: Examples---}
We compare interpolated likelihood functions to those generated by direct evaluation of waveforms and inner products. We consider two test cases: $(i)$ the coalescence of binary black holes, and $(ii)$ the coalescence of binary neutron stars. 

We generate a discretely sampled, simulated data set $\vec{d}$ for a single interferometer consisting of Gaussian and stationary noise $\vec{n}$ and a gravitational-wave signal $\vec{h}$. The data set is $32\,$s in duration and has a sampling rate in the time domain of $4096\,$Hz. For binary black holes we model the gravitational-wave signal $\vec{h}$ using the effective one-body approach calibrated to numerical relativity simulations (EOBNR) \cite{PanEOBNR:2011}. Such a gravitational-wave signal describes the full inspiral, merger and ringdown phases of coalescence. For binary neutron stars we model the gravitational waveform using a post-Newtonian (PN) model computed to 3.5 PN order in phase \cite{Buonanno:2009}, which describes the inspiral phase of the coalescence only. We use an implementation of EOBNR and post-Newtonian waveforms from the LSC Algorithms Library (LAL) \cite{LAL} corresponding to the approximants EOBNRv2 and TaylorT4 respectively. 

%\chad{I guess I hadn't realized in my initial read just how densly the mass space was being sampled. For initial ligo in this mass range there would probably only need to be $\sim$10 templates to cover it, not 625.  I worry now that this is biasing the results.  It is not reasonable to expect tha we could cover the entire parameter space so densly.  In fact, this is so dense that one might be able to do just as well with nearest neighbor interpolation. In other words, simply use the likelihood function from the nearest Mchirp / eta point...  It sounds like you used sbank to create a large template bank (in the practical considerations section) if you were to pick templates from that larger bank around your proposed mass range instead of making the ad hoc 25x25 grid would this still work?}

Generating the interpolant of the likelihood function requires the following stages: $(i)$ patch the mass space into smaller domains, $(ii)$ generate a set of waveforms over a dense grid in each patch, $(iii)$ filter the data with the template waveforms to compute the likelihoods, $(iv)$ pack the likelihoods into a matrix and perform the SVD, $(v)$ build the interpolant in each patch. Only after these stages have been completed can the interpolated likelihood function be sampled.

We first construct a discrete, uniform grid of template waveforms in $\mathcal{M}-\eta$ parameter space. %We fix the number of templates in this space to be $10^4$, which is representative of the number of waveform computations required for the burn-in phase of a MCMC to locate the region in $\mathcal{M}-\eta$ where the posterior has significant support (as contrasted to the $10^6-10^7$ samples required for a complete MCMC).  
We will use a small region around the parameters of the signal, as $\mathcal{M}$ and $\eta$ are typically constrained to $\lesssim 1\%$ and $\lesssim 10\%$, respectively, depending on the  signal parameters and signal-to-noise ratio (SNR) \cite{s6PE, Smith:2012}.  The region in  $\mathcal{M}-\eta$ where the posterior has significant support can be found quickly during the burn-in phase of the MCMC, which requires a small fraction of the total number of samples necessary to evaluate the posterior probability distribution function.

%Our examples are motivated by the fact that the $\mathcal{M}$ and $\eta$ parameters are very well constrained for signals with a reasonable SNR, i.e. for SNR $\sim\,\mathcal{O}(10)$. With such SNRs, $\mathcal{M}$ is typically constrained to within $1\%$, while $\eta$ is typically constrained to within around $10\%$ \cite{s6PE, Smith:2012}. Additionally, for the case of binary neutron stars, $\mathcal{M}$ can be constrained to within around $0.01\%$ and $\eta$ to within around $1\%$ \cite{s6PE}. In practical terms, this means that the MCMC will only explore a relatively small region about the $\mathcal{M}$ and $\eta$ parameters encoded in the signal, after the initial ``burn-in'' phase of the MCMC is complete. Typically, the $2\sigma$ region about the mean parameter estimates is found after around $10^4$ samples, which is to be contrasted with the $10^6-10^7$ samples required for a complete MCMC. Thus, if we could interpolate the likelihood function within the region located by the burn-in of the MCMC, one could would have a novel means to reduce the computational cost of sampling the parameter space for the bulk of the MCMC.}

%, though the maximal match can be close to $100\,\%$. 
 
We use the Chebyshev interpolation described in \cite{Cannon:2012} to interpolate $z[h_0\,,h_0](0)$ for waveforms across the grid. To interpolate $\vec{z}\,[d\,,h_0]$ we first find the basis vectors $\vec{u}_{\mu}$ by constructing a matrix from the set of $\left\{\vec{z}\,[d\,,h_0]\right\}$, the columns of which correspond to a unique $\vec{z}\,[d\,,h_0]$ on the grid of waveforms, which we factor using the SVD. After performing the SVD, we truncate the number of basis vectors such that on average the norm of each $\vec{z}$ is conserved to one part in $10^{5}$ \cite{Cannon:2012}.This can significantly reduce the number of basis vectors.% Imposing this criterion reduces the number of basis vectors and projection coefficients needed to reconstruct $\vec{z}$ from $10^4$ to around 500, which is the total we require for interpolation in $\mathcal{M}-\eta$ space for our two examples.  
We then apply the Chebyshev interpolation \cite{Cannon:2012} to interpolate projection coefficients across the $\mathcal{M}-\eta$ grid.

\textit{Example 1: High-mass binary black holes--}
The signal is parameterized by $\vec{\theta}^{s} = (\mathcal{M} = 15.01\,M_{\odot}\,, \eta = 0.205\,, D=100\,\mathrm{Mpc}\,, \iota=0\,, \psi=0\,, \alpha=0\,,\delta=0\,, t_c=0.1\,\mathrm{s}\,,\phi_{c}=0)$. We use a noise PSD typical of initial LIGO \cite{iLIGO}. The signal has an SNR of $\approx 15$. In order to interpolate the likelihood function across $\mathcal{M}\,, \eta$ and $t_c$, we work within a small region of $\mathcal{M}-\eta$ space whose boundaries are given by $14.56\,M_{\odot}\leq\mathcal{M}\leq15.46\,M_{\odot}$ and $0.143\leq\eta\leq0.25$. Assuming a statistical measurement uncertainty on $\mathcal{M}$ and $\eta$ of $1\%$ and $10\%$, respectively, the parameter ranges correspond to a $\sim3\sigma$ range about the signal value. Note that we cannot go about $\eta=0.25$ in the $\eta$ interval. We further restrict our range in $t_c$ to be in a $\pm0.2\,$s window about the trigger time, which is a common time prior in Bayesian parameter estimation \cite{VV:2010}.

\begin{figure}[htp]
  \centering
  \begin{tabular}{c}
    \includegraphics[scale=0.2]{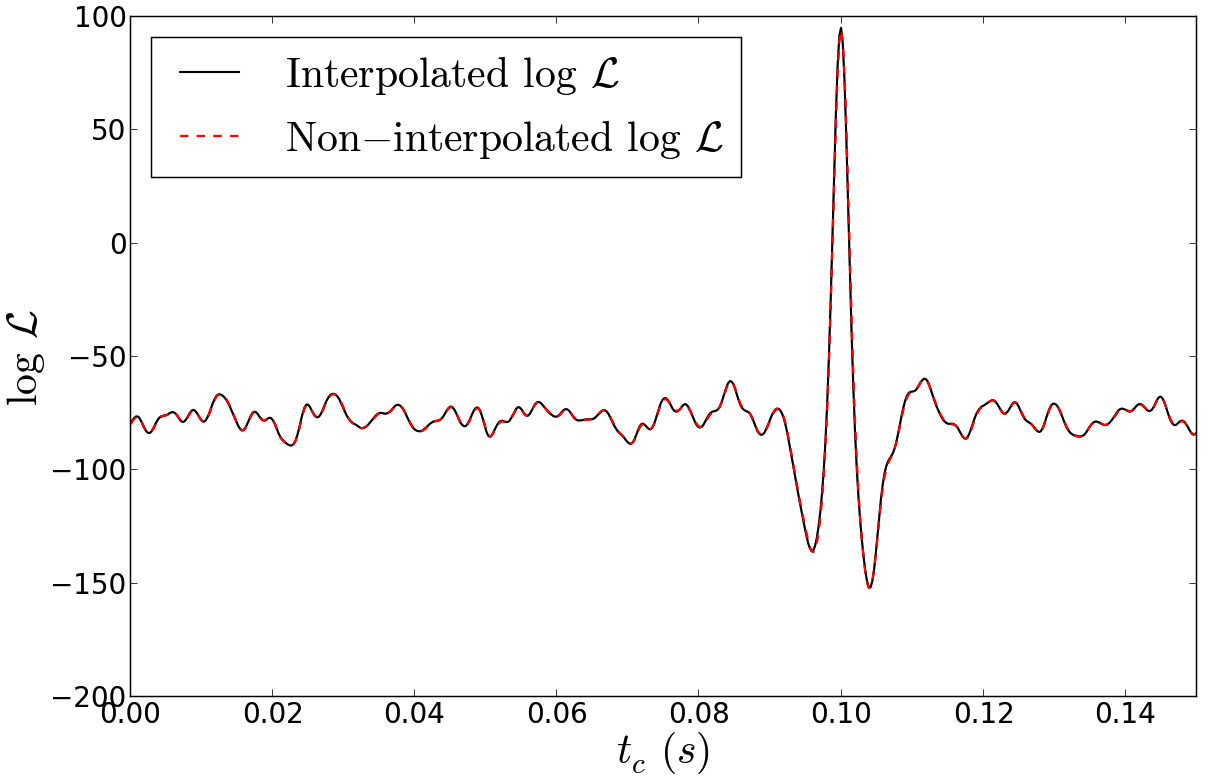}\\
    \includegraphics[scale=0.2]{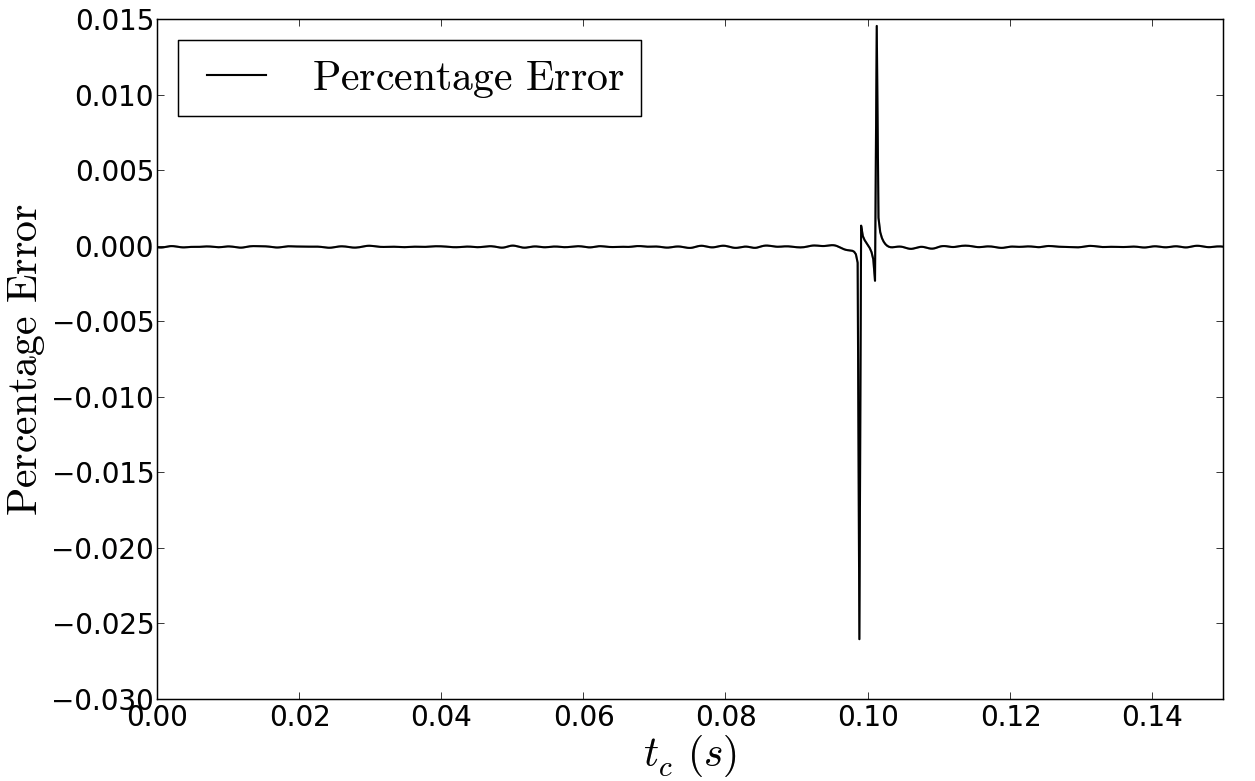}\\
  \end{tabular}
  	  \label{figur}\caption{Interpolated and non-interpolated log likelihoods (top), and percentage error (bottom) for a data set containing a gravitational-wave signal from the coalescence of binary black holes.} 
  	  \label{fig:logL_comp}
\end{figure}

In Fig.~(\ref{fig:logL_comp}) we compare a likelihood function generated via direct evaluation of inner products, to one which we have generated via SVD-interpolation. We find that we are able to reconstruct the log likelihood function by interpolation to within a fractional percentage error of at most $0.025\,\%$. While we have only plotted an interpolated likelihood function at the signal values of $\mathcal{M}$ and $\eta$, the errors quoted here are typical across the mass range we have considered. Meanwhile, for this waveform model and parameters, computing the likelihood via the interpolation procedure is around two orders of magnitude faster than generating a template waveform and directly evaluating the inner products in~(\ref{eq:logL_expanded}).  

\textit{Example 2: Binary neutron stars--}
The signal is parameterized by $\vec{\theta}^{s} = (\mathcal{M} = 1.217\,M_{\odot}\,, \eta = 0.2497\,, D=20\,\mathrm{Mpc}\,, \iota=0\,, \psi=0\,, \alpha=0\,,\delta=0\,, t_c=0.1\,\mathrm{s}\,,\phi_{c}=0)$. We again use a noise PSD typical of initial LIGO \cite{iLIGO}, and the signal has SNR $\approx15$. We interpolate the likelihood function over a small region of $\mathcal{M}-\eta$ space whose boundaries are given by $1.199\,M_{\odot}\leq\mathcal{M}\leq1.235\,M_{\odot}$ and $0.212\leq\eta\leq0.25$. Assuming a statistical measurement uncertainty of $0.5\%$ on $\mathcal{M}$ and $5\%$ on $\eta$, these parameter ranges correspond to a $\sim3\sigma$ range about the signal value. Note that we cannot go about $\eta=0.25$ in the $\eta$ interval. We restrict our range in $t_c$ to be in a $\pm0.2\,$s window about the trigger time.

Again, we find that we are able to reconstruct the log likelihood function to within a fractional percentage error of at most $0.025\,\%$. For the binary neutron star case, we find that
computing the likelihood via interpolation is around three orders of magnitude faster than direct evaluation. This difference is larger than for the higher-mass binary black hole case because the waveform duration is significantly longer for binary neutron stars, whereas the cost of computing interpolated likelihoods remains fixed.

%This illustrates the power of the direct interpolation of the likelihood function.  
Below we discuss practical issues pertaining to incorporating interpolated likelihoods into real gravitational-wave parameter-estimation pipelines.

\textit{Practical considerations---}
For our interpolation technique to be viable for real data analyses, the total computational time of first constructing the interpolated likelihood function, and then sequentially sampling the interpolated likelihood function, must be less than the time for sequentially sampling the likelihood function directly. 
%Crucially, the interpolated likelihood function has to be constructed over an extended region in mass-space typical of a standard parameter estimation analysis. %Incorporating interpolated likelihoods into parameter estimation pipelines requires the following stages: $(i)$ patch up the mass space into reasonable chunks for the SVD $(ii)$ generate all the waveforms required for a given patch, $(iii)$ filter the data with the waveforms in each patch to compute the likelihoods, $(iv)$ pack the likelihoods into a matrix and perform the SVD, $(v)$ build the interpolant in each patch, $(vi)$ sample the interpolated likelihood function. In contrast, directly sampling the likelihood function does not require stages $(i)-(v)$. 
%In Alg.~(\ref{mcmc}) and Alg.~(\ref{hybrid}) we summarize the two approaches, and provide estimates of the computational cost of each in terms of floating point operations (FLOPs).

%\noindent 
Parameter estimation requires sampling the parameter space until the sampler has met its convergence criterion. The total number of likelihood evaluations for convergence is typically $\sim \mathcal{O}(10^{6})$ \cite{Raymond:2009, Smith:2012}.   When directly evaluating the likelihood function, the number of likelihood evaluations is a reasonable proxy for the number of waveform evaluations, which dominate the computational cost. %If the average cost of each waveform evaluation of a given analysis is $\sim\mathcal{O}(T)$, the total cost scales like $\sim \mathcal{O}(T\times 10^{6})$.

To sample the interpolated likelihood function there is an additional upfront cost of constructing the interpolant of the likelihood function. This cost will depend on the region of the parameter space over which the likelihood function needs to be interpolated and template waveforms must be computed.  However, building the interpolant is highly parallelizable and computing it over an extended region of parameter space could be split into multiple independent subsets.  This could greatly reduce the wall time of computing the interpolant.  We have noted that one can restrict the range in parameter space over which the interpolant is built by using an MCMC to sparsely explore the parameter space in regions of high posterior probability. In practice, the number of samples for this ``burn-in'' is often $\sim\,\mathcal{O}(10^{4})$ \cite{Smith:2012}, and the likelihood has significant support in a relatively small patch in parameter space. The likelihoods computed during the burn-in evaluation could thus be stored for future interpolation.

One could also interpolate $\vec{z}(t_c)$ over many patches covering the parameter space in parallel.  We have not investigated optimal patching, nor the required denseness of likelihood template calculation in order to generate a good basis for $\vec{z}(t_c)$; this will be the subject of future work.

Once the interpolant is constructed, the cost of sampling the parameter space will depend on that of computing the interpolated likelihood function. In our example we found that computing the interpolated likelihood function is between two and three orders of magnitude cheaper than directly evaluating the likelihood function, depending on the region in parameter space in which the likelihood function is being computed.  %hence the cost of sampling the interpolated likelihood should scale like $\sim\,\mathcal{O}(T\times\,10^{4})$.   
The actual improvement will depend on the typical cost of waveform computation, which is a function of both the template waveform family used and the waveform parameters.  
%However, we note that this should be considered as an approximate \textit{upper} bound of the FLOPs. 
%We considered short waveforms ($\sim 1\,$s in duration) in our example, but waveforms describing binary neutron star coalescences could last tens of minutes in the advanced detector band.  Thus, the actual cost reduction could significantly exceed the two orders of magnitude quoted above.

%The waveforms in the patch of parameter space in our example are short ($\sim 1\,s$ in duration) and so in areas of the parameter space in which the waveforms are longer, the cost of sampling the interpolated likelihood will be less. For example, waveforms which describe the coalescence of binary neutron stars in aLIGO will be around $10\,$ min in duration and hence sampling an interpolated likelihood here will be several orders of magnitude cheaper than the value quoted in this letter.

%We note that there is a further technicality associated with using interpolated likelihood functions in practice. Parameter estimation analyses typically utilize data sets from multiple interferometers which are combined coherently \cite{}. The coalescence time, $t_c$, of a gravitational-wave signal will be shifted in different interferometers by an amount which corresponds to the light-travel time between them. Thus 

Here we had used the SVD to find a basis for the set of $\vec{z}(t_c)$. The SVD is not a unique technique for finding a basis set, and we note that \citet{Maryland} and \citet{Canizares:2013} employ a greedy algorithm to efficiently generate a set of bases for gravitational waveforms which could in principle be applied to a set of $\vec{z}(t_c)$.

We have so far discussed interpolation in the mass parameters.  It may also be necessary to interpolate the quantities $z[d\,,h_0](t_c)$ in the $t_c$ direction, because the coalescence time in a particular interferometer may lie in between discretely sampled time points.  Second-order interpolation provides sufficient accuracy when the waveform is sampled at $4$ kHz.

\textit{Discussion and conclusion---}
We have demonstrated a method to sample the CBC likelihood function via interpolation, with improvements of two to three orders of magnitude in efficiency. Our method utilizes a SVD of the likelihood function on a three-dimensional subspace of parameters:
%on which the likelihood function has a non-trivial dependence: 
the chirp mass $\mathcal{M}$, symmetric mass-ratio $\eta$ and time at coalescence $t_c$. The SVD factors the likelihood function into a set of scalar coefficients which describe a surface in $\mathcal{M}$ and $\eta$, and a set of orthonormal basis vectors which describe how the surface is translated along $t_c$. The projection coefficients can be interpolated on the $\mathcal{M}-\eta$ plane and then trivially scaled by elements of the basis vectors to generate the likelihood at $(\mathcal{M}\,,\eta\,,t_c)$. This provides an efficient means to interpolate in three dimensions. 

We note that while we have chosen an interpolation technique based on the SVD, it is by no means unique and other interpolation techniques have been applied to gravitational-wave data analysis \cite[e.g.,][]{Canizares:2013}.
%detection statistics have been considered elsewhere \cite{Maryland}. 
Notably, Mitra \textit{et al.} \cite{Mitra:2005} considered interpolating the matched-filtered output of gravitational-wave searches. They interpolated the signal-to-noise ratio, which is 
%output of a matched-filtered search is the signal-to-noise ratio which is the inner product of the data with a template waveform (normalized to unit norm). This quantity is 
effectively a component of the likelihood function, and so their method could, in principle, be extended to interpolate likelihood functions. The key difference with our approach is that we use a decomposition of the likelihood as a function of time, while \cite{Mitra:2005} treat it as a scalar quantity.  This provides us with an efficient means of reducing the total data needed for interpolation, exploiting correlations along the $t_c$ direction by ranking the basis vectors in order of importance in reconstructing the likelihood function. Hence, we can effectively exorcise redundant information based on our accuracy requirements. The number of bases needed to approximately reconstruct the likelihood to high accuracy using the SVD is generally small compared to the number of raw likelihood vectors which we decompose.  %In our example the SVD of a set of $10^4$ likelihood vectors can be described by only 500 basis vectors, thus reducing the computational cost of the interpolated likelihood by a factor of 20.
%and so the amount of data needed to reconstruct the likelihood function is reduced by a factor of 40.

Likelihood interpolation appears to be more robust than waveform interpolation \cite{Smith:2012}, and so utilizing interpolated likelihood functions may also be a stepping stone to tackling the more difficult issue of rapidly estimating the parameters of binaries with spinning components.

\textit{Acknowledgments---} We thank Kipp Cannon and Drew Keppel for many helpful discussions.  RJES acknowledges support from a Perimeter Institute visiting graduate fellowship. Research at Perimeter Institute is supported through Industry Canada and by the Province of Ontario through the Ministry of Research \& Innovation.

\bibliographystyle{apsrev}
\bibliography{./biblio.bib}

\end{document}